\documentclass[11pt,a4paper]{article}
\usepackage{jheppub} 
\usepackage{bm}

\newcommand{\D}{\mathrm{d}}
\newcommand{\mi}{\mathrm{i}}

\title{Torsional response of relativistic fermions in $2+1$ dimensions}

\author{Manuel Valle}

\affiliation{Departamento de F\'\i sica Te\'orica,  
Universidad del Pa\'\i s Vasco UPV/EHU,  
Apartado 644,  48080 Bilbao, Spain}

\emailAdd{manuel.valle@ehu.eus}

\abstract{
We consider the equilibrium partition function of an ideal gas of Dirac fermions 
minimally coupled to torsion in $2+1$ dimensions.   
We show that the energy-momentum tensor reproduces the Hall viscosity 
and other parity violating terms of first order in the torsion.  We also consider 
the modifications of the constitutive relations, and  classify the corresponding susceptibilities. 
An entropy current consistent with zero production of entropy in equilibrium is constructed. 
}

\begin{document}
\maketitle
\flushbottom


\section{Introduction}

The description of transport phenomena and dynamical response 
in the framework of effective field theory has experienced  great progress in the last few years.
This is due, to a large extent, to a better understanding  of the role played by the  
quantum anomalies which underlie new macroscopic 
parity-violating  effects  at low energy~\cite{Son:2009tf}. 
In this respect, the construction of  the thermal partition function as a derivative expansion of 
a time-independent background has turned out to be an important tool to get information about 
the non-dissipative part of the constitutive relations of hydrodynamics, 
without using an entropy approach~\cite{Banerjee:2012iz,Jensen:2012jh}. 
In relativistic systems the first order of such expansion usually has a relatively simple structure
connected with  anomalies~\cite{Jensen:2011xb, Jensen:2012kj, Manes:2013kka}.  
  
For non-relativistic systems there is also considerable interest in establishing 
the precise connection between the partition function and 
Hall transport~\cite{Hoyos:2011ez, Son:2013rqa, Geracie:2014nka, Bradlyn:2014wla, Jensen:2014ama, Gromov:2015sva}.
While in these systems  the use of torsional Newton-Cartan geometry appears as natural, 
the role of torsion in a relativistic setting is less clear~\cite{Haehl:2013kra, Geracie:2014mta}. 
Originally the torsion-dependent effective action of massive Dirac fermions at zero temperature and density  
has been examined in great detail in refs.~\cite{Hughes:2011hv, Hughes:2012vg} 
with the focus on the renormalization effects on the Hall viscosity.
 Here we are interested in the application of the 
 methods of~\cite{Banerjee:2012iz,Jensen:2012jh} to the thermal  
partition function to linear order in the torsion.      

The main purpose of this paper is to compute, to linear order in the torsion,  
the stress tensor and charge current 
that follow from the equilibrium partition function for a Dirac field minimally coupled 
to  torsion.   
A first result of our computation is that the relationship between 
the Hall viscosity and the spin density, 
$\tilde{\eta} = \langle \ell \rangle/2 = -\langle \bar{\Psi} \Psi \rangle/4$,  
naturally appears as an equilibrium susceptibility  
relating the stress tensor $\Theta^{i j}$ with certain parity-violating combination of  
components of the torsion tensor.    
Other results are the explicit construction of an entropy current consistent with zero production of entropy at equilibrium, 
and the derivation of  new susceptibilities relating 
the charge current and the pressure with torsion-dependent
vector and scalar data. 

The organization of the paper is as follows. In section 2  we present some details about the notation, 
and the partition function we use in the subsequent computations. 
The form of the stress tensor at thermal equilibrium  is given in section 3 
in terms of  combinations of scalar, vector, and tensor background data depending on  torsion. 
The modification in the Landau frame of the constitutive relations 
in the presence of torsion is presented in section 4, 
as well as an expression for the entropy current compatible with zero entropy production.  
We conclude in section 5 with a short summary of our results. 

\section{Partition function of a Dirac fermion coupled to torsion} 
We begin with the action for a Dirac field,  
\begin{equation}
\label{eq:dirac}
S =\int \D^3 x \, \text{det}\,  e_\mu^a \left[-\frac{1}{2} \bar{\Psi} \gamma^\mu \overrightarrow{\nabla}_\mu \Psi + 
   \frac{1}{2} \bar{\Psi}  \overleftarrow{\nabla}_\mu \gamma^\mu \Psi   + m \bar{\Psi}  \Psi\right], \qquad 
   \gamma^\mu(x) = e_a^\mu(x) \gamma^a , 
\end{equation}
in the most general
static background  with torsion 
\begin{equation}
\begin{split}
\D s^2 &= G_{\mu\nu} \D x^\mu \D x^\nu = -e^{2\sigma(\bm{x})} (\D t+a_i(\bm{  x}) \D x^i)^2 +g_{ij}(\bm{x}) \D x^i \D x^j , \\ 
\mathcal{A} &= A_0 \D t + \mathcal{A}_i \D x^i , \\ 
T^a &= \frac{1}{2} T_{\mu \nu}^{\;\;\;\, a} \D x^\mu \wedge \D x^\nu .  
\end{split}
\end{equation}
This geometry with torsion has been also considered in other studies of the torsion response~\cite{Hidaka:2012rj}. 
Note that we are using 
the prescription of minimal coupling, where the action depends on the torsion only through 
the covariant derivative. 
The notation of ref.~\cite{sugra} has been adopted,  where  
the Dirac adjoint is defined as $\bar{\Psi} \equiv \Psi^\dagger \mi \gamma^0$. 
For the Dirac matrices $\gamma^a$ we choose the representation 
$\{\gamma^0, \gamma^1, \gamma^2\} = \{ -\mi \sigma_3, \sigma_2, -\sigma_1 \}$. 
The torsion 2-forms $T^a$ are specified in terms of the frame field  $e_\mu^a$ and the torsion tensor 
by $ T_{\mu \nu}^{\;\;\;\, a} = T_{\mu \nu}^{\;\;\;\, \sigma} e_\sigma^a$. 
The covariant derivatives
\begin{equation}
\begin{split}
\overrightarrow{\nabla}_\mu \Psi &= (\partial_\mu + \frac{1}{4} \omega_\mu^{\;\; a  b} \gamma_{a b} - \mi A_\mu ) \Psi , 
           \qquad \gamma_{a b} \equiv \frac{1}{2} [\gamma_a, \gamma_b]  \\
\bar{ \Psi}\, \overleftarrow{\nabla}_\mu &= \bar{ \Psi}  (\overleftarrow{\partial}_\mu - 
\frac{1}{4} \omega_\mu^{\;\; a  b} \gamma_{a b} +\mi  A_\mu ) , 
\end{split}
\end{equation}
are written in terms of the spin connection $\omega_\mu^{\;\; a  b}$, 
which is specified by the 1-forms 
$\omega^{a b} = \omega_\mu^{\;\; a  b} \D x^\mu$ appearing in the Cartan structure equation
\begin{equation}
\label{eq:cartan}
\D e^a +  \omega^{a}_{\; \;\; b} \wedge e^b = T^a . 
\end{equation}
By defining the contortion tensor, antisymmetric in the last two indices,  as 
\begin{equation}
\label{eq:contor}
K_{\mu \nu \rho }= -\frac{1}{2} (T_{[\mu \nu] \rho}  - T_{[\nu \rho] \mu} + T_{[\rho \mu] \nu} ) , 
\end{equation}
it turns out that 
the spin connection can be expressed in terms of the Christoffel symbol 
$\Gamma_{\mu \nu}^{\sigma}(g)$ 
and the contortion tensor as follows 
\begin{equation}
\label{eq:spin}
\begin{split}
\omega_\mu^{\;\; a  b} &= -e^{b \nu} \partial_\mu e_\nu^{a} + 
e^{b \nu}\left(\Gamma_{\mu \nu}^{\sigma}(g) -  K_{\mu \nu}^{\;\;\;\; \sigma}\right)e_\sigma^a  \\ 
&= \omega_\mu^{\;\; a  b}(e) + K_\mu^{\; \; a b} . 
\end{split}
\end{equation}
(The square brackets denote antisymmetrization of indices  $A_{[\mu \nu]}  = \tfrac{1}{2} (A_{\mu \nu} - A_{\nu \mu})$.)
This formula contains the affine connection  
$\Gamma_{\mu \nu}^{\sigma} = \Gamma_{\mu \nu}^{\sigma}(g) -  K_{\mu \nu}^{\;\;\;\, \sigma}$  in 
the presence of torsion, which has an antisymmetric part given by 
\begin{equation}
\label{eq:antis}
\Gamma_{\mu \nu}^{\sigma} - \Gamma_{\nu \mu}^{\sigma} = T_{\mu \nu}^{\;\;\;\, \sigma}. 
\end{equation}
The part of the spin connection in the absence of torsion $\omega_\mu^{\;\; a  b}(e)$, 
which is  called the Levi-Civita connection, is uniquely determined by 
the frame field. It is obtained from (\ref{eq:spin}) by setting $K_{\mu \nu}^{\;\;\;\; \sigma} = 0$. 

In order to vary the metric and torsion variables independently we will assume, according with 
ref.~\cite{Hehl:1976kj}, that 
the torsion tensor with the last coordinate upper index $T_{\mu \nu}^{\;\;\;\, \sigma}$  
and the metric components $G_{\mu \nu}$ 
are independent of each other.  This will be important soon. 
Note that in ref.~\cite{Geracie:2014mta} the contortion is considered as an independent variable, 
instead of the torsion.

With the previous  relations and 
\begin{equation}
\begin{split}
\{\gamma_c, \gamma_{a b} \} &=
 -2 \bar{\epsilon}_{c a b}  , \qquad   \bar{\epsilon}_{0 1 2} = 1 = -\bar{\epsilon}^{0 1 2}, \\
\epsilon^{\mu \nu \rho} &=  \text{det}\,  e_\sigma^d \, 
\bar{\epsilon}^{a b c} e_a^\mu e_b^\nu e_c^\rho ,  
\end{split}
\end{equation} 
we can see the effect of the torsion on the three-dimensional Dirac action. 
The substitution of (\ref{eq:spin}) and (\ref{eq:antis}) into  (\ref{eq:dirac}) reveals that $S$  may be 
viewed as the action of a Dirac field in a torsion-free background, but with a modified mass. 
The new mass $\tilde{m}$, that now depends on the torsion,  is obtained by the replacement
\begin{equation}
\label{eq:effmass}
m \to \tilde{m}  = m + \delta m = m - \frac{1}{8} \epsilon^{\mu \nu \rho} T_{\mu \nu}^{\;\;\;\, \lambda} G_{\lambda \rho} . 
\end{equation}
It is remarkable that a generic non-minimal coupling of the fermion to torsion 
may be included simply by replacing $\delta m$ by $\xi \delta m$,
where $\xi$ is a  free parameter~\cite{Hughes:2012vg}. 

In this way,  the partition function at zeroth derivative order  in the variables  
$(\mathcal{A}_\mu,G_{\mu \nu}, \tilde{m})$ is written as 
\begin{equation}
W^0 = \int \D^2 x \sqrt{g_2} \, \frac{e^\sigma}{T_0} \mathcal{P}(T, \mu, \tilde{m}) , 
\qquad T = T_0 e^{-\sigma}, \qquad \mu = A_0 e^{-\sigma}, 
\end{equation}
where $T_0^{-1}$ is the period of the imaginary time, 
and the function $\mathcal{P}$ is the pressure  in  terms of the temperature and chemical potential.
Therefore,  we expect that the functional  
\begin{equation}
\label{eq:Ztorsion}
\begin{split}
W[G_{\mu \nu}, T_{\mu \nu}^{\;\;\;\, \lambda}]  &= 
-\frac{1}{8} \int \D^2 x \sqrt{g_2} \, \frac{e^\sigma}{T_0} \langle \bar{\Psi} \Psi \rangle \,
 \epsilon^{\mu \nu \rho} T_{\mu \nu}^{\;\;\;\, \lambda} G_{\lambda \rho}  ,
 \end{split}
\end{equation} 
contains all the information about the static linear response  to  lowest order in the derivative expansion of the torsion. 
Here we have used the fact that  the spin density is related to the pressure by 
$\partial \mathcal{P}/\partial m  = \langle \bar{\Psi} \Psi \rangle$, which follows 
from the form of the Dirac Lagrangian, eq.~(\ref{eq:dirac}).  

Although no specific form of $\mathcal{P}$ is required in the following computations,  
we write down  the expressions of the pressure and 
the spin density  at non-zero temperature and density in the free-field case, 
\begin{align}
\mathcal{P}(T, \mu, m) &= -\frac{T^3}{2 \pi} \biggl[ 
\text{Li}_3 \biggl(-\exp\Bigl(\frac{\mu-|m|}{T} \Bigr) \biggr) +
 \text{Li}_3 \biggl(-\exp\Bigl(\frac{-\mu-|m|}{T} \Bigr) \biggr) \biggr] \nonumber  \\ 
 & \quad -\frac{|m| \, T^2}{2 \pi} \biggl[ 
\text{Li}_2 \biggl(-\exp\Bigl(\frac{\mu-|m|}{T} \Bigr) \biggr) +
 \text{Li}_2 \biggl(-\exp\Bigl(\frac{-\mu-|m|}{T} \Bigr) \biggr) \biggr] , \\ 
 \label{eq:spind}
  \langle \bar{\Psi} \Psi \rangle &= -\frac{m T}{2 \pi} \biggl[
  \ln \biggl(1+\exp\Bigl(\frac{\mu-|m|}{T} \Bigr) \biggr) +
  \ln \biggl(1+\exp\Bigl(\frac{-\mu-|m|}{T} \Bigr) \biggr) \biggr] , 
\end{align}
where $\text{Li}_n(x)$ is the polylogarithm function. 
  
\section{The equilibrium stress tensor at linear order in the torsion}
We can now find the energy-momentum tensor $\Theta_{\mu \nu}$ 
which follows from the  partition function 
by differentiation with respect to $\sigma, a_j$ and $g^{i j}$, 
with the understanding that $T_{\mu \nu}^{\;\;\;\, \lambda}$ is independent of the metric. 
From eq.~(\ref{eq:Ztorsion}),  
the variational formula for  $\Theta_{00}$ leads to 
\begin{equation}
\label{eq:t001}
\begin{split}
\Theta_{00} &=  -\frac{T_0  e^\sigma}{\sqrt{g_2}}\frac{\delta W}{\delta \sigma} \\ 
  & = -e^{2 \sigma} \left( \frac{\partial \langle \bar{\Psi} \Psi \rangle}{\partial \sigma} + 
  \langle \bar{\Psi} \Psi \rangle \right) \delta m - 
  e^{2\sigma}  \langle \bar{\Psi} \Psi \rangle \frac{\partial \delta m}{\partial \sigma} .  
\end{split}
\end{equation} 
The most obvious way to obtain $\partial \langle \bar{\Psi} \Psi \rangle/\partial \sigma$ is 
to consider the energy density rewritten in the form 
\begin{equation}
\label{eq:gibbs}
\begin{split}
\varepsilon  &= -\mathcal{P} + T \frac{\partial \mathcal{P}}{\partial T} +  
\mu \frac{\partial \mathcal{P}}{\partial \mu}  \\ 
&= -\mathcal{P} -\frac{\partial \mathcal{P}}{\partial \sigma} .  
\end{split}
\end{equation} 
Then differentiating with respect to $m$ yields the relationship 
\begin{equation}
\frac{\partial  \varepsilon}{\partial m} = -\langle \bar{\Psi} \Psi \rangle - 
\frac{\partial \langle \bar{\Psi} \Psi \rangle}{\partial \sigma}  , 
\end{equation}
and eq.~(\ref{eq:t001})  becomes 
\begin{equation}
\label{eq:T00}
\begin{split}
\Theta_{00} &=  e^{2 \sigma} \frac{\partial  \varepsilon}{\partial m} \delta m - 
  e^{2\sigma}  \langle \bar{\Psi} \Psi \rangle \frac{\partial \delta m}{\partial \sigma} \\ 
 &= e^{2 \sigma} \frac{\partial  \varepsilon}{\partial m} \delta m  + 
 \langle \bar{\Psi} \Psi \rangle \biggl(
 \frac{e^\sigma}{2} \epsilon^{i j} (T_{0 i j} - a_j T_{0 i 0}) -  e^{2 \sigma} \delta m \biggr) . 
\end{split}
\end{equation}
In the computation of the last derivative eq.~(\ref{eq:effmass}) has been used, 
while  keeping $T_{\mu \nu}^{\;\;\;\, \lambda}$ fixed. 
Here $\epsilon^{12} = 1/\sqrt{g_2}$. 
Note the invariance of this result under  time reparametrization, $t \to t + \phi(\bm{x})$,  $\bm{x} \to \bm{x}$. 
In the light of  (\ref{eq:T00}), it is convenient to introduce another  pseudo-scalar quantity (besides $\delta m$) 
constructed from the torsion as  
$\Xi \equiv  -u^\mu T_{\mu \nu \rho} \epsilon^{\nu \rho \lambda} u_\lambda$. 
This quantity,  evaluated for the equilibrium fluid velocity, 
$u_K^\mu = \delta_0^\mu e^{-\sigma}$,  becomes 
$\Xi = -e^{-\sigma} \epsilon^{i j} \mathfrak{T}_{0 i j}$, where $\mathfrak{T}_{0 i j}  = T_{0 i j} - a_j T_{0 i 0}$. 
Thus the change of $\Theta_{0 0}$ induced by the torsion reads
\begin{equation}
\label{eq:T00bis}
\Theta_{0 0} =   e^{2 \sigma} \frac{\partial  \varepsilon}{\partial m} \delta m +  
  e^{2 \sigma} \langle \bar{\Psi} \Psi \rangle 
  \biggl(-\frac{\Xi}{2} -\delta m \biggr) \biggr\rvert_\mathrm{eq}  . 
\end{equation}

The other components  of the energy-momentum tensor are easily computed: 
\begin{equation}
\label{eq:T0i}
\begin{split}
\Theta_0^i &= \frac{T_0 e^{-\sigma}}{\sqrt{g_2}} \frac{\delta W}{\delta a_i} \\ 
  &= \langle \bar{\Psi} \Psi \rangle \biggl(\frac{e^{\sigma}}{4}  \epsilon^{i j}(T_j - a_j T_0)   
   -\frac{ e^{-\sigma}}{2} \epsilon^{i j} T_{0 j 0} \biggr)  ,
\end{split}
\end{equation}
where $T_\mu$ is the torsion vector defined by $T_\mu = T_{\mu \nu}^{\;\;\;\, \nu}$. 
This expression suggests the introduction of  two pseudo-vectors orthogonal to $u^\mu$ given by 
\begin{equation}
\label{eq:seudov}
\begin{split}
\tilde{X}^\mu &= -\epsilon^{\mu \nu \rho} u_\nu u^\lambda T_{\lambda \rho \sigma} u^\sigma ,  \\
\tilde{W}^\mu &= -\epsilon^{\mu \nu \rho} u_\nu T_\rho .   
\end{split}
\end{equation}
In equilibrium these quantities  evaluate  to
\begin{equation}
\begin{split}
\tilde{X}^i &= e^{-2 \sigma} \epsilon^{i j} T_{0 j 0} ,  \\
\tilde{W}^i &=  \epsilon^{i j} (T_j - a_j T_0) . 
\end{split}
\end{equation} 

For the equilibrium stress tensor the functional differentiation of $W$ yields 
\begin{equation}
\label{eq: stress}
\begin{split}
   \Theta^{i j} &= -\frac{2 T_0 e^{-\sigma}}{\sqrt{g_2}} g^{i k} g^{j m} \frac{\delta W}{\delta g^{k m}} \\ 
   &= -\frac{e^{-\sigma}}{4} \langle \bar{\Psi} \Psi \rangle \bigl( \epsilon^{i k} \mathfrak{T}_{0 k m} g^{m j} + 
    \epsilon^{j k} \mathfrak{T}_{0 k m} g^{m i} \bigr) .   
\end{split}
\end{equation}
Remarkably, this expression may be expressed in terms  of the pseudo-tensor given by   
\begin{equation}
\label{eq:dualsigma}
\tilde{\sigma}^{\mu \nu}  = -\frac{1}{2} \bigl(\epsilon^{\mu \lambda \rho} u_\lambda \sigma_{\rho}^{\;\; \nu} + 
 \epsilon^{\nu \lambda \rho} u_\lambda \sigma_{\rho}^{\;\; \mu} \bigr) , 
\end{equation}
where $\sigma^{\mu \nu}$ is the shear tensor, 
\begin{equation}
 \sigma^{\mu \nu} = \Delta^{\mu \alpha} \Delta^{\nu \beta} 
(\nabla_\alpha u_\beta + \nabla_\beta u_\alpha - G_{\alpha \beta} \nabla_\rho u^\rho) , 
\qquad \Delta^{\mu \alpha} = G^{\mu \alpha} + u^\mu u^\alpha. 
\end{equation} 
While in the absence of torsion the equilibrium values of  $\tilde{\sigma}^{\mu \nu}$ and 
$\sigma^{\mu \nu}$ vanish, it turns out that
when torsion is present,  
the covariant derivative of the equilibrium velocity  $u^\mu = \delta_0^\mu e^{-\sigma}$
acquires a non- zero contribution proportional to the torsion, which now is included in the affine connection. 
Correspondingly, the shear tensor at equilibrium is given by 
\begin{equation}
\label{eq:agree}
\begin{split}
 \sigma^{\mu \nu} &= \Delta^{\mu \alpha} \Delta^{\nu \beta} \Bigl[
(K_{\alpha \beta}^{\;\;\;\, \lambda} + K_{\beta \alpha}^{\;\;\;\, \lambda}) u_\lambda + 
G_{\alpha \beta} K_{\nu \lambda}^{\;\;\;\, \nu} u^\lambda \Bigr] \\ 
 &=\Delta^{\mu \alpha} \Delta^{\nu \beta} \Bigl[
-(T_{\lambda \alpha \beta} + T_{\lambda \beta \alpha}) u^\lambda + 
G_{\alpha \beta} T_{\lambda \nu}^{\;\;\;\, \nu} u^\lambda \Bigr] , \qquad u^\lambda = \delta_0^\lambda e^{-\sigma}. 
\end{split}
\end{equation}
Hence, apart from $\tilde{\sigma}^{i j}$ no more pseudo-tensors are needed to express the equilibrium value of $ \Theta^{i j}$.    
Thus  eq.~(\ref{eq: stress}) may be rewritten in the form 
\begin{equation}
\label{eq:hall1}
 \Theta^{i j}  = \frac{1}{4} \langle \bar{\Psi} \Psi \rangle \left( \tilde{\sigma}^{i j}  - \Xi \,  g^{i j} \right) \Bigr\rvert_\mathrm{eq}  .
\end{equation} 

In a general configuration eq.~(\ref{eq:agree}) is not longer satisfied, 
and in  addition to $\tilde{\sigma}^{\mu \nu}$, it is convenient to 
define another pseudo-tensor $\tilde{t}_{(2)}^{\mu \nu}$  related to torsion 
given by  
\begin{equation}
\label{eq:dualt2}
\tilde{t}_{(2)}^{\mu \nu}  = -\frac{1}{2} \bigl(\epsilon^{\mu \lambda \rho} u_\lambda t_{(2)\rho}^{\quad\, \nu} + 
\epsilon^{\nu \lambda \rho} u_\lambda t_{(2)\rho}^{\quad\, \mu} \bigr) , 
\end{equation}
where
 $t_{(2)}^{ \mu \nu}= -\Delta^{\mu \alpha} \Delta^{\nu \beta}(T_{\lambda \alpha \beta} + T_{\lambda \beta \alpha} - 
 G_{\alpha \beta} T_\lambda) u^\lambda$.
This will be important when we consider a possible generalization  to non minimal coupling in the last section, 
 because then $\tilde{t}_{(2)}^{ \mu \nu}$ will appear in the constitutive relations. 
   
It is also possible to derive these results directly from 
the energy-momentum tensor  obtained from the action (\ref{eq:dirac}), without using 
the partition function (\ref{eq:Ztorsion}). 
With the assumption that the spin connection does not depend on the frame field,
the functional derivative  of the action with respect  to $e_a^\rho$ produces the non-symmetric tensor 
\begin{equation}
T'^{\mu \nu} =-\frac{1}{e} e_a^\mu \frac{\delta S}{\delta e_a^{\rho}}  g^{\rho \nu} = 
 \frac{1}{2} 
 \bar{ \Psi} \bigl( \gamma^\mu \overrightarrow{\nabla}^\nu  - 
       \overleftarrow{\nabla}^\nu \gamma^\mu  \bigr) \Psi  .
\end{equation}
The symmetry may be restored by adding the tensor $\Delta T^{\mu \nu}$ given by 
\begin{equation}
\Delta T^{\mu \nu} = \frac{1}{8}  (\nabla_\lambda + 
   T_\lambda ) \bigl( \bar{\Psi} \{\gamma^{\lambda \mu}, \gamma^\nu \} \Psi \bigr) , 
\end{equation}
where the covariant derivative is performed with the affine connection including the contortion,    
and $T_\lambda$ denotes  the torsion vector $T_\lambda \equiv T_{\mu \rho}^{\;\;\;\, \rho}$. 
Using the equations of motion 
this gives the usual form of the stress tensor  
\begin{equation}
\label{eq:sym}
\Theta^{\mu \nu} = T'^{\mu \nu} + \Delta T^{\mu \nu} = \frac{1}{4} 
 \bar{ \Psi} \bigl( \gamma^\mu \overrightarrow{\nabla}^\nu  - 
       \overleftarrow{\nabla}^\nu \gamma^\mu  + (\mu \leftrightarrow \nu) \bigr) \Psi .
\end{equation}
To obtain the dependence of $\langle \Theta^{\mu \nu} \rangle$ with the torsion,  we note that 
the contortion tensor in the covariant derivatives always  multiplies the  
term $\langle \bar{\Psi} \Psi \rangle$.
A second kind of contribution comes from the fact that the fermion field satisfies a Dirac equation 
with a torsion-dependent mass $\tilde{m}$. As a consequence the corresponding perfect fluid constitutive relation 
\begin{equation}
\langle \Theta^{\mu \nu} \rangle= \varepsilon(T, \mu, \tilde{m}) u^\mu u^\nu + 
\mathcal{P}(T, \mu, \tilde{m}) \Delta^{\mu  \nu} , 
\end{equation}  
contributes to the expectation values of  $\Theta_{0 0} $ and  $\Theta^{i j}$ 
with terms proportional to $\partial \varepsilon/\partial m$ and 
$\partial \mathcal{P}/\partial m $ respectively. 
Then the explicit evaluation of $K_\mu^{\; \; a b}$ in the spin connection 
shows that the equilibrium values may be written as the combinations 
\begin{align}
\label{eq:stress00}
\langle\Theta_{0 0}\rangle &=     
 \langle \bar{\Psi} \Psi \rangle \biggl(
 \frac{e^\sigma}{2} \epsilon^{i j} \mathfrak{T}_{0 i j}  -  e^{2 \sigma} \delta m \biggr) + 
 e^{2 \sigma} \frac{\partial  \varepsilon}{\partial m} \delta m , \\ 
 \label{eq:stress2}
\langle\Theta^{i j}\rangle &=  
-\langle \bar{\Psi} \Psi \rangle \biggl(\frac{e^{-\sigma}}{4} \bigl( \epsilon^{i k} \mathfrak{T}_{0 k m} g^{m j} + 
    \epsilon^{j k} \mathfrak{T}_{0 k m} g^{m i} \bigr)  +  \delta m \,  g^{i j} \biggr) + 
     \frac{\partial \mathcal{P}}{\partial m}  \delta m \,  g^{i j} . 
\end{align} 
These are in perfect agreement with (\ref{eq:T00})  and (\ref{eq: stress}).
The result for $\langle \Theta_{0}^j \rangle$  coincides with~(\ref{eq:T0i}), 
and  does not include contributions of the second type because they are not generated by 
the perfect fluid constitutive relation. 

\subsection{Kubo formula for Hall viscosity from a contact term} 
So far we have been working at the level of the partition function, but it is instructive to consider 
the response to a general time-dependent applied torsion in the framework of linear response theory.
With this in view,  we now consider 
eq.~(\ref{eq:stress2}) as the static limit of a Kubo formula
for the stress tensor response to an externally applied time-dependent torsion (not necessarily uniform).
In order to obtain such a formula, 
we assume that the only non-zero components of the torsion tensor are $T_{0 j k} =-T_{j 0 k}$, 
and we also consider that  the metric is Minkowskian. 
Thus the change of the Hamiltonian  to linear order in the torsion becomes  
\begin{equation}
 \begin{split}
 H_1 &= -\int \D^2 x  \,  \bar{\Psi}  \Psi \delta m \\ 
  &= -\frac{1}{4} \int \D^2 x  \,  \bar{\Psi}  \Psi \,   \epsilon^{i j} T_{0 i j} .  
 \end{split}
 \end{equation}
The induced change of the expectation value of $\Theta_{i j}$ is  obtained from linear response theory 
 in the form 
\begin{equation}
\label{eq:lrt}
\begin{split}
\delta \langle \Theta_{i j}(\bm{x}, t) \rangle &= \int \D^2 y 
\left\langle \frac{\delta \Theta_{i j}(\bm{x}, t)}{\delta T_{0 k m}(\bm{y},t)} \right\rangle_0 T_{0 k m}(\bm{y},t) \\
& \quad +
\frac{i}{4}  \int_{-\infty}^t \D \bar{t}\,  e^{-\eta (t-\bar{t})} \int \D^2 y  \left \langle [\Theta_{i j}(\bm{x}, t),  \bar{\Psi}  \Psi(\bm{y},\bar{t}) ] \right\rangle_0
\epsilon^{k m} T_{0 k m}(\bm{y},\bar{t}), 
\end{split}
\end{equation}
where $\eta \to 0^+$, and 
the subindex $0$ means an equilibrium average with respect to the unperturbed Hamiltonian.  
The first term (contact term) has its origin in the explicit dependence of $\Theta$ with the torsion, 
and corresponds exactly to the first summand  of eq.~(\ref{eq:stress2}), 
\begin{equation}
\left\langle \frac{\delta \Theta_{i j}(\bm{x}, t)}{\delta T_{0 k m}(\bm{y},t)} \right\rangle_0 = 
-\frac{1}{4} \langle \bar{\Psi}  \Psi \rangle_0 \left(
\epsilon^{i  k} \delta^{m j} + \epsilon^{j  k} \delta^{m i} + \epsilon^{k m} \delta^{i j} \right) 
\delta(\bm{x}-\bm{y})  \, .
\end{equation}
The second term involves the retarded stress tensor-spin density correlator defined by  
\begin{equation}
\mathcal{Y}_{i j}(\bm{x},t) \equiv -\frac{i}{2} \lim_{\eta \to 0^+}  \theta(t)  
 \left \langle [\Theta_{i j}(\bm{x}, t),  \bar{\Psi}  \Psi(\bm{0},0) ] \right\rangle_0 e^{-\eta t} .  
\end{equation}
We can express eq.~(\ref{eq:lrt}) in terms of a 
response function to the torsion  $\chi_{i j\, k m}(\bm{q}, \omega)$, which in the Fourier domain relates
  $\delta \langle \theta_{i j} \rangle = 
\chi_{i j\, k m} T_{0 k m}$.   
Therefore eq.~(\ref{eq:lrt})  leads to the following Kubo-type formula
\begin{equation}
\label{eq:kubo}
\chi_{i j\, k m}(\bm{q}, \omega) = -\frac{1}{4} \langle \bar{\Psi}  \Psi \rangle_0 \left(
\epsilon^{i  k} \delta^{m j} + \epsilon^{j  k} \delta^{m i} \right)  - \frac{1}{2} 
\Bigl( Y_{i j}(\bm{q},\omega)  + \frac{1}{2} \langle \bar{\Psi}  \Psi \rangle_0\delta_{i j} \Bigr)
\epsilon^{k m} , 
\end{equation}
where 
\begin{equation}
Y_{i j}(\bm{q},\omega) = \int_0^\infty \D t \, e^{i (\omega + i 0^+) t} \int \D^2 x \,   e^{-i \bm{q} \cdot \bm{x}}  
\left \langle [\Theta_{i j}(\bm{x}, t),  \bar{\Psi}  \Psi(\bm{0},0) ] \right\rangle_0 . 
\end{equation}
The response function $Y_{i j}(\bm{q},\omega)$ 
is similar to the (integrated) stress-strain form of the response function  
in  ref.~\cite{Bradlyn:2012fk},
where the role  of the spin density is played by the antisymmetric part of the strain generators $J_{\alpha \beta}$.

We see that the Kubo formula~(\ref{eq:kubo}) is consistent with the static result of eq.~(\ref{eq:stress2}) 
only if the ordered limit 
$\lim_{q\to 0} \lim_{\omega \to 0} Y_{i j} (\bm{q}, \omega)$ satisfies 
\begin{equation}
\lim_{q\to 0} \lim_{\omega \to 0} Y_{i j} (\bm{q}, \omega) = \lim_{q\to 0}Y_{i j} (\bm{q}, \omega=0) = 
-\frac{\delta_{i j} }{2} \frac{\partial \mathcal{P}}{\partial m}. 
\end{equation}
This requirement can be expressed as a thermodynamic sum rule  valid for $q \to 0$, 
\begin{equation}
\lim_{q \to 0} \int_{-\infty}^\infty \frac{\D \omega}{\pi} \frac{\text{Im} Y_{i j} (\bm{q},\omega) }{\omega-i0^+} =   
-\frac{\delta_{i j}}{2} \frac{\partial \mathcal{P}}{\partial m}  , 
\end{equation}
which is similar to other sum rules giving different susceptibilities. 
It follows that the static response to $T_{0 k m}$ is determined by a part of the contact term, namely
\begin{equation}
\lim_{q\to 0} \chi_{i j\, k m}(\bm{q}, \omega=0) = 
\tilde{\eta} \left(
\epsilon^{i  k} \delta^{m j} + \epsilon^{j  k} \delta^{m i} \right) ,
\end{equation}
so that 
\begin{equation}
\tilde{\eta} = \frac{1}{4} \lim_{q \to 0} \delta_{i j} \epsilon_{k m} \chi_{i j\, k m}(\bm{q}, \omega=0) .
\end{equation}  
The coefficient $\tilde{\eta}=-\langle \bar{\Psi}  \Psi \rangle_0/4$  will be identified below with the Hall viscosity.  


\section{Torsion-dependent constitutive relations}

We now consider the corrections to the constitutive relations 
imposed by the partition function (\ref{eq:Ztorsion}). 
The argument essentially follows ref.~\cite{Banerjee:2012iz}. 
As these corrections involve the expectation value of the particle current $J^\mu = \mi \,  \bar{\Psi} \gamma^\mu \Psi$, 
we  also need  
the modification induced by the torsion on the equilibrium current.  
This  is simply given by 
\begin{equation}
\label{eq:dj0}
\delta J_0 =  -\frac{T_0  e^\sigma}{\sqrt{g_2}}\frac{\delta W}{\delta A_0} 
= -e^\sigma \frac{\partial n}{\partial m} \delta m, \qquad \delta J^i = 0.  
\end{equation}
This may be immediately obtained from the relation between the particle density and the pressure, 
$n= e^{\sigma} \partial \mathcal{P}/\partial A_0$, 
which implies 
\begin{equation}
\frac{\partial \langle \bar{\Psi} \Psi \rangle}{\partial A_0} = 
e^{-\sigma} \frac{\partial n}{\partial m} . 
\end{equation}

\subsection{Stress tensor and charge current}
The most general parity violating modification of the currents  that 
includes torsion must be written in terms of the non-zero quantities in equilibrium 
 $\delta m, \Xi, \tilde{W}^\mu, \tilde{X}^\mu$ and $\tilde{\eta}^{\mu \nu}$.  
 Then  the non-dissipative parts of the constitutive relations in the Landau frame take the form 
 \begin{equation}
 \begin{split}
\label{eq:pimunu}
  T^{\mu \nu} &= \varepsilon u^\mu u^\nu + \mathcal{P} \Delta^{\mu \nu}  -\left(\tilde{\chi}_\Upsilon \delta m + 
 \tilde{\chi}_\Xi \Xi\right) \Delta^{\mu \nu}  - \tilde{\eta}\, \tilde{\sigma}^{\mu \nu}  , \\
 J^\mu &= n u^\mu + \tilde{ \chi}_W \tilde{W}^\mu + \tilde{ \chi}_X \tilde{X}^\mu , 
\end{split}
\end{equation}
which involve four new transport coefficients $\tilde{\chi}_\Upsilon, \tilde{\chi}_\Xi, 
 \tilde{ \chi}_W, \tilde{ \chi}_X$ and the Hall viscosity $\tilde{\eta}$.\footnote{
 The term $\left(\tilde{\chi}_\Upsilon \delta m + 
 \tilde{\chi}_\Xi \Xi\right) \Delta^{\mu \nu}$ would be replaced by 
   $\left(\tilde{\chi}_B B + \tilde{\chi}_ \Omega \Omega + \tilde{\chi}_\Upsilon \delta m + 
 \tilde{\chi}_\Xi \Xi\right) \Delta^{\mu \nu}$  in the complete constitutive relation, 
 where $B = \tfrac{1}{2} \epsilon^{\mu \nu \rho} F_{\nu \rho}$ and the vorticity is defined by 
$\Omega =  \epsilon^{\mu \nu \rho} u_\mu \partial_\nu u_\rho$ rather than  
$\epsilon^{\mu \nu \rho} u_\mu \nabla_\nu u_\rho$. 
Then the effect of the additional term  related 
to the antisymmetric part of the affine connection in the presence of  torsion, 
$\Delta \Omega = -\epsilon^{\mu \nu \rho} u_\mu T_{\nu \rho} u^\lambda$, 
is considered separately in the coefficients 
 $\tilde{\chi}_\Upsilon$ and $\tilde{\chi}_\Xi$.}  
To find them,  we specialize to equilibrium 
 by using $u^\mu = \delta_0^\mu e^{-\sigma}$ to zero order in the torsion, and 
$\Delta_{00}= \tilde{\sigma}_{00} = \tilde{X}_0 =\tilde{W}_0 = 0$.  
Because  the torsion-induced corrections  can be expressed
in terms of changes in the fluid variables $(\delta T, \delta \mu, \delta u^i)$  according to    
\begin{equation}
\label{eq:chang}
\begin{split}
\delta T_{0 0}& =e^{-2 \sigma}\left( \frac{\partial \varepsilon}{\partial T} \delta T  + 
 \frac{\partial \varepsilon}{\partial \mu} \delta \mu \right) ,  \\ 
 \delta J_{0}& = -e^{-\sigma}\left( \frac{\partial n}{\partial T} \delta T  + 
 \frac{\partial n}{\partial \mu} \delta \mu \right) ,  \\ 
 \delta T_0^i  &=-e^\sigma (\varepsilon + \mathcal{P}) \delta u^i, \\
  \delta J^i  &= n \,  \delta u^i  +\tilde{ \chi}_W \tilde{W}^i + \tilde{ \chi}_X \tilde{X}^i, \\ 
 \delta T^{i j} &= \left( \frac{\partial \mathcal{P}}{\partial T} \delta T  + 
 \frac{\partial \mathcal{P}}{\partial \mu} \delta \mu -\tilde{\chi}_\Upsilon \delta m - 
 \tilde{\chi}_\Xi \Xi \right)  g^{i j} - \tilde{\eta}\, \tilde{\sigma}^{i j} , 
\end{split}
\end{equation}
we can equate  these first three equations 
to (\ref{eq:T00bis}), (\ref{eq:dj0}) and  (\ref{eq:T0i}). This gives  
\begin{equation}
\label{eq:deltas}
\begin{split}
\left(\frac{\partial \varepsilon}{\partial T} \frac{\partial n}{\partial \mu} - 
 \frac{\partial \varepsilon}{\partial \mu} \frac{\partial n}{\partial T} \right) \delta T &= \left[
 \frac{\partial n}{\partial \mu} \left(\frac{\partial \varepsilon}{\partial m} -  \langle \bar{\Psi} \Psi \rangle \right) - 
  \frac{\partial \varepsilon}{\partial \mu} \frac{\partial n}{\partial m} \right] \delta m - 
 \frac{ \langle \bar{\Psi} \Psi \rangle}{2} \frac{\partial n}{\partial \mu} \Xi , \\ 
 \left(\frac{\partial \varepsilon}{\partial T} \frac{\partial n}{\partial \mu} - 
 \frac{\partial \varepsilon}{\partial \mu} \frac{\partial n}{\partial T} \right) \delta \mu &= \left[
 \frac{\partial n}{\partial T} \left(-\frac{\partial \varepsilon}{\partial m} +  \langle \bar{\Psi} \Psi \rangle \right) + 
  \frac{\partial \varepsilon}{\partial T} \frac{\partial n}{\partial m} \right] \delta m + 
 \frac{ \langle \bar{\Psi} \Psi \rangle}{2} \frac{\partial n}{\partial T} \Xi , \\ 
 \delta u^i & = \frac{\langle \bar{\Psi} \Psi \rangle}{\varepsilon + \mathcal{P}}\left( 
 \frac{\tilde{X}^i}{2} - \frac{\tilde{W}^i}{4} \right) . 
 \end{split}
\end{equation}
We now determine the transport coefficients appearing in the constitutive relations 
by equating the last two equations in (\ref{eq:chang}) with  (\ref{eq:dj0}) and (\ref{eq:hall1}).  
With eqs.~(\ref{eq:deltas}) and the thermodynamical derivatives 
\begin{equation}
\begin{split}
\frac{\partial \mathcal{P}}{\partial \varepsilon} &= 
 \left(\frac{\partial \mathcal{P}}{\partial T} \frac{\partial n}{\partial \mu} - 
 \frac{\partial \mathcal{P}}{\partial \mu} \frac{\partial n}{\partial T}\right)\biggl/
  \left(\frac{\partial \varepsilon}{\partial T} \frac{\partial n}{\partial \mu} - 
 \frac{\partial \varepsilon}{\partial \mu} \frac{\partial n}{\partial T}\right) , \\ 
 \frac{\partial \mathcal{P}}{\partial n} &= 
 \left(-\frac{\partial \mathcal{P}}{\partial T} \frac{\partial \varepsilon}{\partial \mu} + 
 \frac{\partial \mathcal{P}}{\partial \mu} \frac{\partial \varepsilon}{\partial T}\right)\biggl/
  \left(\frac{\partial \varepsilon}{\partial T} \frac{\partial n}{\partial \mu} - 
 \frac{\partial \varepsilon}{\partial \mu} \frac{\partial n}{\partial T}\right)  , 
 \end{split}
\end{equation}
we are left with 
\begin{equation}
\label{eq:suscep}
\begin{split}
\tilde{\chi}_\Upsilon &= \langle \bar{\Psi} \Psi \rangle  \left(1-\frac{\partial \mathcal{P}}{\partial \varepsilon}  \right)   ,  \\
\tilde{\chi}_\Xi &= -\langle \bar{\Psi} \Psi \rangle  \left(\frac{1}{2} \frac{\partial \mathcal{P}}{\partial \varepsilon}  +
\frac{1}{4} \right)   ,  \\
\tilde{\eta} &= -\frac{\langle \bar{\Psi} \Psi \rangle}{4} ,   \\
\tilde{ \chi}_W &= \frac{1}{4} \langle \bar{\Psi} \Psi \rangle \frac{n}{\varepsilon + \mathcal{P}} , \\ 
\tilde{ \chi}_X &= -\frac{1}{2} \langle \bar{\Psi} \Psi \rangle \frac{n}{\varepsilon + \mathcal{P}} . 
\end{split}
\end{equation}
These expressions, which uniquely determine the new transport coefficients in term of the angular momentum density,  
 are one of the main results of this paper. 

 
 \subsection{The entropy current}
Finally we consider the contribution to the entropy current which arises 
from  the partition function in eq.~(\ref{eq:Ztorsion}), 
\begin{equation}
\label{eq:Ztorsion2}
W = \int \D^2 x \sqrt{g_2} \, \frac{e^\sigma}{T_0} 
\frac{\partial \mathcal{P}(T, \mu,m)}{\partial m} \delta m  , \qquad 
\delta m  = -\frac{1}{8}  \epsilon^{\mu \nu \rho} T_{\mu \nu \rho} .  
\end{equation}
If $s$ denotes the entropy density that follows from the thermodynamic potential at zero derivative order, 
$s = \partial \mathcal{P}/\partial T$,  
the standard formula for the entropy yields the contribution 
 \begin{equation}
S = \frac{\partial (T_0 W)}{\partial T_0} = \int \D^2 x \sqrt{g_2} \,  \frac{\partial^2 \mathcal{P}}{\partial m\,  \partial T} \delta m = 
\int \D^2 x \sqrt{g_2} \,  \frac{\partial s}{\partial m} \delta m.   
\end{equation}
Therefore a condition that must be fulfilled by the correction to the 
entropy current  $\delta J_S^\mu$ is 
\begin{equation}
\int \D^2 x \sqrt{g_2} \, e^\sigma \delta J_S^0=\int \D^2 x \sqrt{g_2} \,  
\frac{\partial s}{\partial m} \delta m . 
\end{equation}  
Let us consider the following tentative expression for the entropy current containing the 
dependence on the torsion:
\begin{equation}
J_S^\mu = s u^\mu - \frac{\mu}{T} \left(\tilde{ \chi}_W \tilde{W}^\mu + \tilde{ \chi}_X \tilde{X}^\mu \right) + (b_\Xi \, \Xi  + b_\Upsilon \delta m) u^\mu + 
 d_W \tilde{W}^\mu + d_X \tilde{X}^\mu . 
\end{equation} 
It follows then from this form of the entropy current that the four coefficients 
$b_\Xi$, $b_\Upsilon$, $d_W$ and $d_X$ are determined, in a unique way, by the condition  
\begin{equation}
\begin{split}
\delta J_S^0 &=  s \,  \delta u^0 + e^{-\sigma} \left(\frac{\partial s}{\partial T} \delta T + \frac{\partial s}{\partial \mu} \delta \mu \right) \\
&\quad - \frac{\mu}{T} \left(\tilde{ \chi}_W \tilde{W}^\mu + \tilde{ \chi}_X \tilde{X}^\mu \right) + (b_\Xi \, \Xi  + b_\Upsilon \delta m) u^\mu + 
 d_W \tilde{W}^\mu + d_X \tilde{X}^\mu  \\ 
 &= e^{-\sigma} \frac{\partial s}{\partial m} \delta m . 
\end{split}
\end{equation}
Then one finds that
\begin{equation}
\label{eq:ventro}
\begin{split}
b_\Xi &= \frac{\langle \bar{\Psi} \Psi \rangle}{2 T}, \\
b_\Upsilon &= \frac{2 \langle \bar{\Psi} \Psi \rangle}{T} , \\ 
d_W &= \frac{\langle \bar{\Psi} \Psi \rangle}{4 T}, \\
d_X &= -\frac{\langle \bar{\Psi} \Psi \rangle}{2 T} . 
\end{split}
\end{equation}
Note that in order to obtain these results it is necessary to use the values of $\delta T$ and $\delta \mu$ given by  
 eqs.~(\ref{eq:deltas}), as well as the expressions for 
 $\delta u^0 = -a_i \delta u^i$, $\tilde{W}^0$, $\tilde{X}^0$,  and the 
 thermodynamical identities 
\begin{equation}
\frac{\partial s}{\partial \varepsilon} =  \frac{1}{T}, \qquad \frac{\partial s}{\partial n} =  -\frac{\mu}{T}, 
\qquad \varepsilon + \mathcal{P} = T s + \mu n.  
\end{equation}
Thus one has
\begin{equation}
\frac{\partial s}{\partial m} = \frac{1}{T}  \frac{\partial \varepsilon}{\partial m} - 
\frac{\mu}{T}  \frac{\partial n}{\partial m}. 
\end{equation}

It is remarkable that the results in eqs.~(\ref{eq:ventro}) imply that 
the spatial part of $\delta J_S^\mu$  is identically zero at equilibrium: 
\begin{equation}
\begin{split}
\delta J_S^i &=  s \,  \delta u^i  + 
\left(-\frac{\mu}{T} \tilde{\chi}_W + d_W \right) \tilde{W}^i + 
\left(-\frac{\mu}{T} \tilde{\chi}_X + d_X \right) \tilde{X}^i \\ 
&=  s \,  \delta u^i  + 
\frac{s \langle \bar{\Psi} \Psi \rangle}{\varepsilon + \mathcal{P}}\left( 
 \frac{\tilde{W}^i}{4} - \frac{\tilde{X}^i}{2} \right) \\ 
 &=0 . 
\end{split}  
\end{equation}
Thus the condition of no entropy production 
is satisfied since the divergence of $\delta J_S^\mu$  is necessarily zero under these conditions.\footnote{If torsion is present, 
the covariant differentiation is defined with respect to the non-symmetric connection  
$\Gamma_{\mu \nu}^{\sigma}(g) - K_{\mu \nu}^{\;\;\;\, \sigma}$. However, it is possible to define a 
modified divergence (see e.g., ref.~\cite{Hehl:1976kj}) according to  
$\overset{*}\nabla_\mu \equiv \nabla_\mu + T_{\mu \nu}^{\;\;\;\, \nu}$, 
which for a vector field $V^\mu$ reduces to the torsion-free result 
$\overset{*}\nabla_\mu V^\mu =( \sqrt{-g})^{-1} \partial_\mu (\sqrt{-g} V^\mu)$. 
Thus the conservation or the entropy current found above must correspond to $\overset{*}\nabla_\mu J_S^\mu = 0$.}

\section{Conclusion}

In this paper, we have studied the modifications of the stress tensor  
of an ideal gas of Dirac fermions 
coupled minimally to torsion. 
We have concentrated on the part coming from the equilibrium thermal 
partition function to linear order in the torsion, 
which in $2+1$ dimensions may be easily expressed in terms of an effective mass shift. 
Differentiating with respect to the metric we have obtained different types of background data linear in the torsion, including two pseudo-scalars, two pseudo-vectors and a pseudo-tensor field.  
We have also found an expression for the entropy current consistent  with zero entropy production at equilibrium. 
It turns out that the non-dissipative part of the constitutive relations involve five susceptibilities, 
four of them expressing the modifications  of  the stress tensor and the current in response to the 
applied torsion.  
Because of the way the torsion affects the covariant derivative of the fluid velocity, 
the remaining susceptibility may be written as the proportionality factor that 
expresses the equilibrium response to the parity odd shear tensor.  
Therefore it would be identified as the Hall viscosity.

Throughout  this paper we have assumed that the fermions are minimally coupled to torsion. 
Considering non-minimal couplings introduces an additional parameter on which some of response functions 
may depend.  Specifically,  one has to make the  replacement $\delta m \to \xi \delta m$ for arbitrary $\xi$, 
therefore leading at first sight  to $\tilde{\eta} \to \xi \tilde{\eta}$ .  
This has been used to question the existence of a connection between Hall viscosity and 
torsion response~\cite{Geracie:2014mta}.
However, the analysis sketched above ignores the fact, pointed out in  section 3,   
that there are actually two independent tensors, given by eqs.~(\ref{eq:dualsigma}) and (\ref{eq:dualt2}), 
which  coincide at equilibrium (see eq.~(\ref{eq:agree})). 
Therefore, both tensors must be used in the modified constitutive relation   
\begin{equation}
\delta T_{\mu \nu} =  -\tilde{\eta}\, \tilde{\sigma}_{\mu \nu}  -  \tilde{\eta}_2 \,  \tilde{t}_{(2)}^{\mu \nu}. 
\end{equation}
Now, as  the covariant derivative in the presence of torsion involves the 
affine connection  $\Gamma_{\mu \nu}^{\sigma} = \Gamma_{\mu \nu}^{\sigma}(g) - K_{\mu \nu}^{\;\;\;\, \sigma}$,  
rather than $\Gamma_{\mu \nu}^{\sigma}(g) -  \xi K_{\mu \nu}^{\;\;\;\, \sigma}$, 
the parity odd shear tensor $\tilde{\sigma}^{\mu \nu}$ in the last term of eq.~(\ref{eq:pimunu}) 
remains unchanged even if  $\xi \neq 1$, 
and the same applies to $\tilde{\eta}$ which cannot depend on $\xi$. 
Therefore the torsional response in the presence of non-minimal coupling requires that 
\begin{equation}
\tilde{\eta}_2 = (1-\xi)\frac{\langle \bar{\Psi} \Psi \rangle}{4} . 
\end{equation}
In this way the Hall viscosity remains intact, and in the absence of torsion is the only non-dissipative term  
in the constitutive relation. 

Finally, it would be desirable to combine the results  from holographic models with those presented 
here, but it does not appear to be an easy task.
The main reason is that  there is no unique way to produce Hall viscosity 
and angular momentum density in 
holography~\cite{Saremi:2012uq, Jensen:2011xb, Son:2014fk, Wu:2013vya}.   
For example,  the holographic model considered in ref.~\cite{Jensen:2011xb} gives rise to  
nonzero angular momentum density but vanishing Hall viscosity, 
while for the model considered in ref.~\cite{Son:2014fk} the ratio between these quantities 
is compatible with the universal value 1/2 from field theory. 
In this respect, it may be interesting to note that for a gapped system,  such as a free massive Dirac fermion,  the spin density of eq.~(\ref{eq:spind}) is nonzero at $\mu=0$, while the angular momentum density $\langle \ell \rangle$ from 
certain holographic models~\cite{Jensen:2011xb}   vanishes at  $\mu=0$.  Thus   
the relation $\langle \ell \rangle = \tfrac{1}{2} \partial \mathcal{P}/\partial m$ that  determines many features of the results in this paper  
does not seem to be satisfied in those models. 
Further studies are required to understand all these points. 


\acknowledgments
I would like to thank Juan L. Ma\~nes for useful discussions and remarks. 
This work has been supported by Plan Nacional de Altas Energ\'\i as FPA2012-34456, Consolider Ingenio 2010 CPAN CSD2007-00042 and by the Basque Government under Grant IT559-10.

\bibliographystyle{JHEP}
\bibliography{bibtorsion}

\end{document}